\documentclass{article}
\usepackage{amsmath}
\usepackage{graphicx}
\usepackage{amsfonts}
\usepackage{amssymb}
\begin{document}

\begin{center}
{\LARGE Entanglement and Bell Inequalities.}

\textbf{M.Kupczynski\medskip}

\ Department \ of Mathematics and Statistics , Ottawa University\medskip
\end{center}

\begin{quotation}
\textbf{Abstract.} The entangled quantum states play a key role in quantum
information. The association of the quantum state vector with each individual
physical system in an attributive way is a source of many paradoxes and
inconsistencies. The paradoxes are avoided if the purely statistical
interpretation (SI) of the quantum state vector is adopted. According to the
SI the quantum theory (QT) does not provide any deterministic prediction for
any individual experimental result obtained for a free physical system, for a
trapped ion or for a quantum dot. In this article it is shown that if the SI
is used then, contrary to the general belief, the QT does not predict for the
ideal spin singlet state perfect anti-correlation of the coincidence counts
for the distant detectors. Subsequently the various proofs of the Bell's
theorem are reanalyzed and in particular the importance and the implications
of the use of the unique probability space in these proofs are elucidated. The
use of the unique probability space is shown to be equivalent to the use of
the joint probability distributions for the non commuting observables. The
experimental violation of the Bell's inequalities proves that the naive
realistic particle like spatio- temporal description of the various quantum
mechanical experiments is impossible. Of course it does not give any argument
for the action at the distance and it does not provide the proof of the
completeness of the QM. The fact that the quantum state vector is not an
attribute of a single quantum system and that the quantum observables are
contextual has to be taken\ properly into account in any implementation of the
quantum computing device.

\bigskip

\textbf{Keywords:}Entanglement , Bell's inequalities, quantum information,
quantum computing , \ EPR correlations, quantum cryptography
\end{quotation}

\textbf{PACS Numbers}: 03.65. Bz, 03.67. -a, 03.67. Dd, 03.67.Hk

\subsection{Introduction}

\noindent The long range non classical correlations characterizing the
entangled quantum states are at the base of the quantum computer project[42,
32, 26], state teleportation and quantum cryptography [24, 9, 17]. The
mathematical structure and possible time evolutions of the quantum states have
been studied and a considerable progress has been achieved [29,4,47]. The
entanglement witnesses have been constructed which may help to distinguish
between different entangled states in the experiment [40, 18]. Quantum states
and quantum process tomography have been studied and experimentally
implemented [41,42,13,30,31] . In spite of this incontestable progress of
quantum information in some papers the state vectors (qubits) are treated as
the attributes of the individual quantum system which can be manipulated and
modified quasi- instantaneously. One may also occasionally find the picture of
the Schr\"{o}dinger cat and hear a story of the twin point-like particles
communicating at the distance with faster than light signals. It seems that
the abstract, statistical and contextual character of the quantum description
of the Nature is sometimes forgotten. Besides it is usually assumed that a
single measurement reduces instantaneously the state vector of a physical system.

The problems related to the quantum theory of the measurement and a notion of
the state vector reduction have been for decades a subject of discussions
between people interested in the foundations of the quantum theory (QT) and
still there is no unanimity. The most consistent seems to us a point of view
of the followers of so called purely statistical interpretation (SI) of QT
which evolved from the interpretation advocated for the first time by
Einstein.[23,22] . According to SI the pure state vector $\Psi$ or the density
matrix $\varrho$ describes only the statistical properties of an ensemble of a
similarly prepared systems. For the trapped ions and the quantum dots it
describes the statistical properties of the repeated measurements on the same
ion or the quantum dot after the same initial preparation. The statistical
interpretation was extensively discussed by Ballentine $\left[  14\right]  $.
In his already classic textbook of the quantum mechanics based on the SI we
may read $\left[  15\right]  $: ''Once acquired , the habit of considering an
individual particle to have its own wave function is hard to break. Even
though it has been demonstrated to be strictly incorrect, it is surprising how
seldom it leads to a serious error.'' \ In the SI the state vector reduction
is a passage from the description of the whole ensemble to the description of
the sub-ensemble obtained from the initial ensemble by so called non
destructive measurements.The important additional arguments in favour of the
SI have been recently given by Allaverdyan, Balian and Nieuwenhuizen$\left[
48\right]  .$

Since most of the predictions of the QT are of statistical nature a famous EPR
question [23] might be asked whether and in what sense the QT provides a
complete description of the individual physical system. In fact the SI leaves
in principle a place for the introduction of the supplementary parameters
(called often hidden variables) which would determine the behavior of each
particular physical system during the experiment. Several theories with
supplementary parameters (TSP) have been discussed $\left[  7\right]  $. The
most influential was the paper by Bell$\left[  8\right]  ,$who analyzed a
large family of TSP so called local or realistic hidden variable theories
(LRHV) and showed that their predictions must violate,\ for some
configurations of the experimental set-up, the quantum mechanical predictions
for spin polarization correlations experiments(SPCE) dealing with pairs of
electrons or photons produced in a singlet state . Bell's argument was put
into experimentally verifiable form, by Clauser, Horne,Shimony and
Holt$\left[  19\right]  $. Several experiments in particular those by Aspect
et al. $\left[  5,6\right]  $ confirmed the predictions of QM. The general
conclusion summarized in the excellent review by Clauser and Shimony $\left[
21\right]  $ was that if one wants to understand the experimental data ''
either one must totally abandon the realistic philosophy of most working
scientists or dramatically revise our concept of space time '' which
encouraged unwillingly speculations about a spooky action on a distance.

It was shown by many authors that assumptions made in LRHV were more
restrictive and questionable that they seemed to be and the Bell's
inequalities may be violated not only by quantum experiments but also by
macroscopic ones[1,3, 35-38].The recent experiments seemed to close the
remaining loop holes [45,43 ] but the violation of CHSH\ may not be consider
neither as a proof of the completeness of QM nor the indication of the faster
than light communication [44, 29, 2, 30, 39, 25] . The extensive discussion of
the concept of \ probability were given by Khrennikov[49] and Holevo[50]. The
role of the contextuality and the remaining loopholes in Bell's proof were
recently underlined by Khrennikov and Volovich [51-53].

In this short paper we want to refine and complement some of our old arguments
and forgotten ideas [35-38] hoping that it could shed some light on the
problems we face in the quantum information.

The paper is organized as follows in the section 2 we reanalyze\ in view of
the SI the properties of the entangled idealized spin singlet state. In
particular we show that there is no prediction for the perfect correlations of
the counts for the far away detectors and no EPR-Bohm paradox. Let us
underline that this lack of perfect correlations is of more deep nature than
the lack of perfect correlations in all real experiments which is attributed
to the decoherence, experimental systematic and statistical errors and
efficiency of detectors.\{21,30,25]. In section 3 we analyze some proofs of
Bell and CHSH inequalities clearly demonstrating that the use of the unique
probability space is equivalent to the use of the joint probability
distributions for the noncommuting observables or to the assumption that all
random variables corresponding to physical observables studied are completely
independent thus uncorrelated. Let us note that most of the proofs of the
recent generalizations of the CHSH inequalities to the n qubits are usually
done assuming the factorization of the expectation functions thus the
statistical independence of the corresponding random variables.

\subsection{{\protect\LARGE A singlet state.}}

Let us state the{\LARGE \ }essential points of the EPR -Bohm \ reasoning using
the notation and phrasing from the reference $\left[  15\right]  .$

The singlet spin state vector for the system of two particles has the form

$\Psi_{0}$ = $\left(  \mid+\text{ }\rangle\otimes\text{ }\mid-\text{ }%
\rangle\text{ - }\mid-\text{ }\rangle\otimes\text{ }\mid+\text{ }%
\rangle\right)  \sqrt{1/2}$ \ \ \ \ \ \ \ \ \ \ \ \ \ \ \ \ \ \ \ \ \ \ \ \ \ \ \ \ \ \ \ \ \ \ \ \ \ \ \ \ \ \ \ \ \ \ \ \ \ \ \ \ \ \ \ (1)\ \ \ \ \ \ \ \ \ \ \ \ \ \ \ \ \ \ \ \ \ \ \ \ \ \ \ \ \ \ \ \ \ \ \ \ \ \ \ \ \ \ \ \ \ \ \ \ \ \ \ \ \ \ \ \ \ \ \ \ 

where the single particle vectors $\mid+$ $\rangle$ and $\mid-$ $\rangle$
denote ''spin up'' and ''spin down '' with respect to some coordinate system.

a) Even if the orbital state is not stationary, the interactions do not
involve spin and so the spin states will not change.

b) The particles are allowed to separate , and when they are well beyond the
range of the interaction we can measure the z component of spin of the
particle \#1.

c) Because the total spin is zero, we can predict with certainty, and without
in any way disturbing the second particle, that the z component of spin of
particle \#2 must have the opposite value. Thus the values of $\sigma
_{z}^{\left(  2\right)  }$ is an element of reality , according to EPR \ criterion.

d) But the singlet state is invariant under rotation and it has the same form
(1) in term of ''spin up'' and ''spin down'' if the directions ''up'' and
''down'' are referred to any other axis. Thus following EPR we may argue that
the values of $\sigma_{x}^{\left(  2\right)  }$, $\sigma_{y}^{\left(
2\right)  }$ and any number of other spin components are also elements of the
reality for the particle \#2.

What is wrong with this argument? In a) all possible decoherence due to the
interaction with the environment is neglected . In b) by saying that the
particles had a time to separate we assume a mental image of two point-like
particles which are produced and which after some time become separated and
free. Even if we assume that the points a) and b) are correct then the point
c) is wrong and it will be proven below using the SI. We do not see any
particular couple of the particles and we do not follow its space time
evolution. We record only the clicks on the far away coincidence counters. To
be able to deduce the value of a particular spin projection for the particle
\#2 from the measurement made on the particle \#1 we should have had for each
experiment (A,B) \ a different experimental design ( impossible to realize)
giving us much more information \ about each couple of the particles than we
have in a simple coincidence experiment. Similar arguments were given by Bohr
$\left[  12,11\right]  $ in his neither well understood nor frequently read
answer to the original EPR paper. We interpret a click as a detection of the
particle which passed by a polarization filter and which was registered by a
detector. According to SI only\ the ensemble of these particles is described
by the one particle state vector $|+$ $\rangle$ or $|-$ $\rangle$ with respect
to the axis determined by the filter. Let us note that if c) is not correct
than d) does not follow and there is no EPR\ paradox. According to SI a state
$\Psi_{0}$ allows only to find the statistical correlations observed in a long
run of the various experiments with different couples (A,B) of the spin
polarization analyzers, characterized by macroscopic direction vectors
\textbf{A} and\textbf{\ B. }Since the angle between \textbf{\ A} and
\textbf{B} is a continuous variable the QT gives us the probability density
functions not the probabilities. Let's go back to the mathematical formalism
of the QT .

Let $\ \sigma_{\mathbf{a}\text{ \ \ }}=$ \textbf{\ }$\mathbf{\sigma}\bullet$
\textbf{a} \ \ denote the component of the Pauli spin operator in the
direction of the unit vector \textbf{a} , and $\sigma_{\mathbf{b}\text{ \ \ }%
}=$ \textbf{\ }$\mathbf{\sigma}\bullet$ \textbf{b} \ \ denote the component of
the Pauli spin operator in the direction of the unit vector \textbf{b} . If we
''measure'' the spin of the particle \#1 along the direction \textbf{a} and
the spin of particle \#2 along the direction\textbf{\ b} , the results will be
correlated, and for the singlet state the correlation is

$\left\langle \Psi_{0}\left|  \sigma_{\mathbf{a}\text{ \ \ }}\otimes\text{
}\sigma_{\mathbf{b}\text{ \ \ }}\right|  \Psi_{0}\right\rangle $ = - cos
$\theta_{\mathbf{ab}}$
\ \ \ \ \ \ \ \ \ \ \ \ \ \ \ \ \ \ \ \ \ \ \ \ \ \ \ \ \ \ \ \ \ \ \ \ \ \ \ \ \ \ \ \ \ \ \ \ \ \ \ \ \ \ \ \ \ \ $\ \ \ \ \ \ \ \ \ \ \ \ \ \ \ \ \ \ \left(
2\right)  $

where $\theta_{\mathbf{ab}}$ is the angle between the directions \textbf{a}
and\textbf{\ b}.

Each spin polarization correlation experiment (A,B) is defined by two
macroscopic orientation vectors\ \textbf{A} \ and\textbf{\ B} being some
average orientation vectors of the analyzers. An analyzer \textit{\ A} is
defined by a probability distribution d$\rho_{A}(\mathbf{a})$ , where
\textbf{a }are the microscopic direction vectors, \textbf{a}$\in
O_{A}=\left\{  \mathbf{a}\in S^{(2)};\left|  1-\mathbf{a}\cdot\mathbf{A}%
\right|  \leq\varepsilon_{A}\right\}  .$ Similarly a polarizer B is defined by
d$\rho_{B}(\mathbf{b}).$\ The probability p(A,B) that a \ particle \#1 is
detected by the analyzer A and a particle \#2 , correlated with the particle
\#1 is detected by a analyzer B could be given by

p(A,B)= $\ \eta\left(  A\right)  \eta\left(  B\right)  \ \ \underset{O_{A}%
}{\int}$ $\ \underset{O_{B}}{\int}p_{12}(\mathbf{a},\mathbf{b})$d$\rho
_{A}(\mathbf{a})$d$\rho_{B}(\mathbf{b})$\ \ \ \ \ \ \ \ \ \ \ \ \ \ \ \ \ \ \ \ \ \ \ \ \ \ \ \ \ \ \ \ \ \ \ \ \ \ \ \ \ \ \ (3)

where $p_{12}(\mathbf{a},\mathbf{b})$ is a probability density\textbf{\ }%
function given by QM :

$p_{12}(\mathbf{a},\mathbf{b})$= $\frac{1}{2}\sin^{2}(\theta_{ab}/2)$ $and$
$\eta$'s are some factors related to the efficiency of the detectors.
Similarly the predicted correlation function$\ E(A,B)$ to be compared with the
experimental data is given by

\bigskip E(A,B)= $\ \eta\left(  A\right)  \underset{}{\eta\left(  B\right)
\text{ }\underset{O_{A}}{\int}\ \underset{O_{B}}{\int}}\underset{}%
{-cos\theta_{\mathbf{ab}}\ }$d$\rho_{A}(\mathbf{a})$d$\rho(\mathbf{b})$
\ \ \ \ \ \ \ \ \ $\ \ \ \ \ \ \ \ \ \ \ \ \ \ \ \ \ \ \ \ \ \ \ \ \ \ \ \ \ \ \ \ \ \ \left(
4\right)  $

We see that the observable value of the spin projection characterizes only the
whole beam of the'' particles'' which passed through a given \ analyzer A.
Nearly 100\% of the '' particles'' of this beam would pass by the subsequent
identical analyzer A, but we have no prediction concerning any individual
''particle'' from the beam. and we have no strict spin anti-correlations
between the members of \ each pair.

Let us now discuss the various proofs of the Bell's inequalities.\bigskip

\subsection{Bell's Theorem\ }

For any random experiment we may find a non unique mathematical probabilistic
model describing it . Given a probabilistic model there exist in general
several random experiments which can be described by the model. To obtain the
consistency of the probabilistic model with the experiment a particular
experimental design and a protocol have to be adopted. It was clearly
demonstrated by Bertrand $\left[  10\right]  $ and discussed by us $\left[
38,39\right]  .$

To each random experiment we associate a random variable X, a probability
space S and a probability density function f$_{X}$(x) for all x$\in$ S.

{\Large \ }If\ X is a discrete random variable $\underset{x}{\sum}$ f$_{X}
$(x)=1 and P(X=x)=f$_{X}$(x) \ If X is a continuous random variable$\underset
{S}{\int}$\ f$_{X}$(x)dx=1 and

\bigskip

P(a$\leq$X$\leq$b)\ =$\underset{a}{\int}$\ $^{b}$ f$_{X}$(x)dx \ \ \ \ \ \ \ \ \ \ \ \ \ \ \ \ \ \ \ \ \ \ \ \ \ \ \ \ \ \ \ \ \ \ \ \ \ \ \ \ \ \ \ \ \ \ \ \ \ \ \ \ \ \ \ \ \ \ \ \ \ \ \ \ \ \ \ \ \ \ \ \ \ \ \ \ \ \ \ \ \ \ \ \ \ \ \ \ \ (5)

\bigskip

where P(a$\leq$X$\leq$b) is a probability of finding a value of X included
between a and b. Note that P(X=x) = 0 for all x$\in$S\textbf{.}

\bigskip

If in a random experiment we can measure simultaneously values of k- random
variables\ X$_{1}$,...X$_{k}$ we describe the experiment by a k-dimensional
random variable X= (X$_{1}$,..,X$_{k}$), \ a common probability space S and
some joint probability density function f$_{X_{1}X_{2}..X_{k}}$ (x$_{1}%
$,..x$_{k}$) . From the joint probability density function we can obtain
various conditional probabilities and by integration over k-1 variables we
obtain k marginal probability density functions f$_{X_{i}}$(x$_{i}$)
describing \textit{k} different random experiments each performed to measure
only one random variable X$_{i}$ and neglecting all the others . In this case
we say that f$_{X_{i}}$ (x$_{i\text{ }}$) were obtained by conditionalization
from a unique probability space S. In general if the random variables X$_{i}$
are dependent (correlated)

\bigskip

f$_{X_{1}X_{2}..X_{k}}$ (x$_{1}$,..x$_{k}$)$\neq$f$_{X_{1}}$(x$_{1}$)
f$_{X_{2}}$(x$_{_{2}}$)...f$_{X_{k}}$(x$_{k}$) \ \ \ \ \ \ \ \ \ \ \ \ \ \ \ \ \ \ \ \ \ \ \ \ \ \ \ \ \ \ \ \ \ \ \ \ \ \ \ \ \ \ \ \ \ \ \ \ \ \ \ \ \ \ \ \ \ \ \ \ \ (6)

\bigskip

As we found in the preceding section each spin polarization correlation
experiment (A,B) is defined by two macroscopic orientation vectors\ \textbf{A}
\ and\textbf{\ B} and the coincidence probabilities are given by (3) and the
correlation functions are given by (4). It is impossible to perform different
experiments (A,B) simultaneously on the same couple of the particles therefore
it does not seem possible to use a unique probability space S and to obtain,
by conditionalization, the probabilities p(A,B) for all such experiments. This
is why that it is not so strange that Bell's inequalities proven using a
common probability space do not agree with the predictions of QT.

\bigskip

Let us now analyze a model used by Clauser and Horne $\left[  20\right]  $to
prove their inequalities:

\bigskip

p(A,B)=$\underset{\Lambda}{\int}p_{1}(\lambda,A)$ $p_{2}(\lambda
,B)d\rho(\lambda)$ \ \ \ \ \ \ \ \ \ \ \ \ \ \ \ \ \ \ \ \ \ \ \ \ \ \ \ \ \ \ \ \ \ \ \ \ \ \ \ \ \ \ \ \ \ \ \ \ \ \ \ \ \ \ \ \ \ \ \ \ \ \ \ \ \ \ \ \ \ \ (7)\ 

\bigskip

where $p_{1}(\lambda,A)$ and $p_{2}(\lambda,B)$ are the probabilities of
detecting component 1 and component 2 respectively , given the state $\lambda$
of the composite system .

We see from (7) that a state $\lambda$ is determined by all the values of
strictly correlated spin projections of two components for all possible
orientations of the polarizers A and B. The polarizers are not perfect
therefore the detection probabilities have been introduced. Therefore it is
assumed in the model that even before the detection each component has well
defined spin projection in all directions. The model is using a single
probability space $\Lambda$ and obtains the predictions on the probabilities
p(A,B) measured in different experiments by conditionalization. As we told the
same assumption was used in all other proofs of Bell's theorem. Explicit
description of states $\lambda$ by the values of spin projections is also
clearly seen in Wigner's proof$\left[  46\right]  $.As we told the experiments
(A,B) are mutually exclusive so there is no justification for using such models.

\bigskip

If we try to prove the Bell's inequalities by comparing only the experimental
runs of different experiments we can not do it without some additional and
questionable assumptions.

Let us simplify the argument we gave in$\left[  38\right]  .$We want to
estimate a value of the spin expectation function E$\left(  A,B\right)  $ for
an experiment (A,B) . We have to perform several runs of the length N and find
the value of the empirical spin expectation function r$_{N}$(A,B) for each run
and after to estimate E$\left(  A,B\right)  $ by averaging over various runs.
Let us associate with each member of a pair a spin function s$_{1}$(x) or
s$_{2}$(x), taking the values 1 or -1,\ on the unit sphere S$^{(2)}$
(representing the orientation vectors of various polarizers) .We assume also that

s$_{1}$(x) =- s$_{2}$(x)= s(x) for all vectors x$\in$S$^{(2)}$. We saw in
equation (3) that the macroscopic directions \textbf{A} and\ \textbf{B} were
not sharp therefore in each particular run we might have different direction
vectors (\textbf{a},\textbf{b}) representing them. If for the simplicity we
neglect this possibility, we get:\bigskip

r$_{N}$(A,B)= $\ -\frac{1\text{ \ }}{N}\underset{i}{\sum}$ s$_{i}$%
(\textbf{A})s$_{i}$(\textbf{B}) \ \ \ \ \ \ \ \ \ \ \ \ \ \ \ \ \ \ \ \ \ \ \ \ \ \ \ \ \ \ \ \ \ \ \ \ \ \ \ \ \ \ \ \ \ \ \ \ \ \ \ \ \ \ \ \ \ \ \ \ \ \ \ \ \ \ \ \ \ \ \ \ \ \ (8)

where N functions s$_{i}$ are drawn from some uncountable set of \ \ spin
functions F$_{0}.$

If we consider a particular run of the same length from the experiment (A,C)
we get

\bigskip r$_{N}$(A,C)= $\ -\frac{1\text{ \ }}{N}\underset{j}{\sum}$ s'$_{j}%
$(\textbf{A})s'$_{j}$(\textbf{C}) \ \ \ \ \ \ \ \ \ \ \ \ \ \ \ \ \ \ \ \ \ \ \ \ \ \ \ \ \ \ \ \ \ \ \ \ \ \ \ \ \ \ \ \ \ \ \ \ \ \ \ \ \ \ \ \ \ \ \ \ \ \ \ \ \ \ \ \ \ \ \ \ \ (9)

where N functions s'$_{j}$ are drawn from the same uncountable set of \ spin
functions F$_{0}.$

A probability that we have the same sets of spin functions in both
experimental runs is equal to zero. Therefore in general we have completely
distinct sets of functions in (8) and (9) and we are unable to prove the
Bell's theorem by using r$_{N}$(A,B)-r$_{N}$(A,C). If we used the same sets of
spin functions in the runs from the different experiments then we could
replace (9) by (10)

\bigskip r$_{N}$(A,C)=$-\frac{1\text{ \ }}{N}\underset{i}{\sum}$ s$_{i}%
$(\textbf{A})s$_{i}$(\textbf{C}) \ \ \ \ \ \ \ \ \ \ \ \ \ \ \ \ \ \ \ \ \ \ \ \ \ \ \ \ \ \ \ \ \ \ \ \ \ \ \ \ \ \ \ \ \ \ \ \ \ \ \ \ \ \ \ \ \ \ \ \ \ \ \ \ \ \ \ \ \ \ \ \ \ \ \ \ (10)

and we could easily reproduce the Bell's proof \ finding \ his inequalities in
the standard form or in the form given for the first time in the reference
$\left[  15\right]  $ :

\bigskip$\left|  E(A,B)-E(A,B^{\prime})\right|  +\left|  E(A^{\prime
},B^{\prime})+E(A^{\prime},B)\right|  \leq$ 2 \ \ \ \ \ \ \ \ \ \ \ \ \ \ \ \ \ \ \ \ \ \ \ \ \ \ \ \ \ \ \ \ \ \ \ \ \ \ \ \ \ \ \ \ \ \ (11)

One could still have some doubts concerning the above argument for the sharp
directions of the polarizers.( the samples are not the same but on the long
run everything should average out, etc.) However if the directions of the
polarizers are not sharp our random experiment is not only a random sampling
from some unique population of the spin functions followed by their exact
evaluation. In the subquantal description of the experiment (A,B) we have 3
populations: population of couples of correlated spin functions, microscopic
directions of the polarizer A and microscopic directions of the polarizer
B.The sampling from these three populations produce the effective samples of
the experimental data which are sets of couples of the numbers $\pm
1$\ corresponding to a draw from these populatios and the evaluation of the
spin functions.. Therefore if we change the experiment into (C,D) the results
may not be represented by conditionalization from some unique probability
space common for (A,B) and C,D). The smearing of the polarization directions
is important in the impossbility of the rigorous proof of the Bell
inequalities in this type of subquantal description of the phenomenon.

When the validity of the inequality is tested (11) one should estimate
properly all the quantities and \ include the correct error bars [25].

Let us also notice that the act of passage of the i-th particle through a
given analyzer\ A depends in a complicated way on its interaction with this
polarizer. Therefore we should not consider a spin function as describing a
state of a particle independent of its interaction with A. The spin functions
s$_{i}$ in the (8) and (9) resume the interactions of the subsequent particles
with the polarizers in a particular experiment. Therefore if we want to be
rigorous we should replace (8) by (12).

r$_{N}$(A,B)= $\ -\frac{1\text{ \ }}{N}\underset{i}{\sum}$ s$_{i,\mathbf{A}}%
$(\textbf{a}$_{i}$)s$_{i,\mathbf{B}}$(\textbf{b}$_{i}$) \ \ \ \ \ \ \ \ \ \ \ \ \ \ \ \ \ \ \ \ \ \ \ \ \ \ \ \ \ \ \ \ \ \ \ \ \ \ \ \ \ \ \ \ \ \ \ \ \ \ \ \ \ \ \ \ \ \ \ \ \ \ \ \ \ \ (12)

where \textbf{a}$_{i}\in O_{A}$ and \textbf{b}$_{i}\in O_{B}.$ \ If we use the
formula (12) there is no possibility of proving Bell's theorem . Using this
formula we can always obtain the results consistent with the equation (3) .The
formula (12) visualizes the contextual character of the observables.

In a trivial but artificial way a common probability space S could be used in
a case if we had four\textbf{\ }independent experiments\textbf{\ \ }%
described\textbf{\ } by four independent random variables X$_{1}$,X'$_{1}%
,$X$_{2}$,X'$_{2}$ and their probability density functions. If all possible
values of these variables had the absolute value smaller or equal to 1 a proof
of Bell's inequalities would be extremely easy . In such a case the ''spin''
expectation function E(X$_{1},$X$_{2}$) is a product of expectation values of
X$_{1}$ and X$_{2}$ : E(X$_{1},$X$_{2}$)$=\left\langle X_{1}\right\rangle
\left\langle X_{2}\right\rangle $ and we immediately get

$\bigskip$

$\left|  \left\langle X_{1}\right\rangle \left\langle X_{2}\right\rangle
-\left\langle X_{1}\right\rangle \left\langle X_{2}^{\prime}\right\rangle
\right|  $ + $\left|  \left\langle X_{1}^{\prime}\right\rangle \left\langle
X_{2}^{\prime}\right\rangle +\left\langle X_{1}^{\prime}\right\rangle
\left\langle X_{2}\right\rangle \right|  \leq$

$\bigskip$

$\leq\left|  \left\langle X_{2}\right\rangle -\left\langle X_{2}^{\prime
}\right\rangle \right|  +\left|  \left\langle X_{2}\right\rangle +\left\langle
X_{2}^{\prime}\right\rangle \right|  \leq2$ \ \ \ \ \ \ \ \ \ \ \ \ \ \ \ \ \ \ \ \ \ \ \ \ \ \ \ \ \ \ \ \ \ \ \ \ \ \ \ \ \ \ \ \ \ \ \ \ \ \ \ \ \ \ \ \ \ \ \ (13)

\bigskip

which is the exactly the inequality (11)

Of course if we assume the independence there are no correlations. The
statistical independence is related \ to the separability of the statistical
operator used recently by Kr\"{u}ger in his proofs of Bell's inequalities in
$\left[  33\right]  $ .

In the similar way the quantum correlations are neglected in the cryptographic
proof by Herbert$\left[  29\right]  $ reviewed by Ballentine $\left[
15\right]  $ . The source of the singlet state is represented as a generator
of the two correlated signals. If the two detectors (A,B) are aligned in the
same directions the two messages ( the strings of +1and -1 are identical). If
the detector B is rotated by an angle $\theta$ it is assumed that the rate of
disagreement between the two d($\theta)$ is due only to the change in the
orientation of B and does not depend on the orientation of the spatially
\ separated detector A. This assumptions leads to the inequality
d(2$\theta)\leq$ 2d($\theta)$ which does not agree with the predictions of QM.
Let us note that quantum mechanical correlations are the correlations between
the counts of the distant detectors obtained by the coincidence technique and
they are never perfect..The messages, string of the bits, are not send by the
source they are only created by the coincidence technique after the results of
the measurements for each pair of the analyzers (A,B) are recorded. Therefore
in each experiment the rate of the disagreement depends on the directions of
both macroscopic devices not only on the one of them. Before the measurement
there is no message. In Herbert's approach the rate of disagreement is treated
like a measure of the random errors of reading some preexisting incoming
message which depends on the rotation of only one of the analyzers from its
initial position. The subtle quantum mechanical statistical correlations
between counts of A and B are simply ignored and the contextual character of
the quantum observables is neglected.

\subsection{Conclusions{\protect\LARGE .}}

The violation of the Bell inequalities requires neither the abandon of the
Einstenian separability nor the abandon of the realistic point of view
according to which external reality is assumed to exist and to have definite
properties. The properties of the reality are however not attributive but
contextual. Without doubt in the SPCE a source is producing the pulses of some
real physical field. These pulses are interacting with far away analyzers and
produce the correlated clicks of the detectors. The interference and the
diffraction of light has been successfully explained by the wave picture of
Huygens and Maxwell in the classical physics. The violation of the Bell
inequalities forces us to abandon naive realistic models according to which
the source is producing a stream of couples of point like particles flying to
the detectors , the couples having well defined individuality and the
properties possessed in the attributive way. The subquantal intuitive picture,
if it did exist, it would have to be of a completely different nature. This
subquantal picture is however not needed. Quantum theory with its statistical
interpretation provides the algorithms allowing to explain the results of the
experiments in the microworld without providing any spatio- temporal
description of the physical phenomena involved.

The lack of the deterministic predictions for the individual measurements and
the SI interpretation of the quantum state vectors have implications for the
quantum information. There is no problem with the implementation of quantum
cryptography since transmission of the secret key can be realized successfully
with the use of the short pulses of polarized light or with gaussian-
modulated coherent states $\left[  32\right]  $ instead of using the single photons.

The fact that the quantum state vector is not an attribute of a single quantum
system requires more caution in the problems related to the implementation of
the quantum computing devices $\left[  42,32,26\right]  .$ A more detailed
discussion of the contextual character of quantum observables [11,34,39] and
its implication for the quantum computing will be given in the following paper.\bigskip

\subsection{References}

\begin{enumerate}
\item  Accardi,L, Phys.Rep.\textbf{77}(1981), 169.

\item  Accardi, L.and Regoli, M,, Locality and Bell's
Inequality,in:A.Khrennikov(ed.) QP-XIII, Foundations of Probability and
Physics, , World Scientific , Singapore,2002, 1--28

\item  Aerts,D., J.Math.Phys. \textbf{27}(1986),202

\item  Alicki, R.and Fannes, M, Quantum Dynamical Systems, Oxford Univ. press,
Oxford, 2001.

\item  Aspect,A., Grangier P.and Roger,G., Phys.Rev.Lett.\textbf{47} (1981)
460; \textbf{49}(1981) 91:

\item  Aspect,A., Dalibard, J.and Roger,G.,Phys. Rev.Lett. \textbf{49} (1982), 1804.

\item  Belinfante,F.J., A Survey of Hidden Variable Theories,Pergamon, New
York, 1973.

\item  Bell, J.S.,Physics \textbf{1}(1965)195..

\item  Bennet,H., Brassard G., and Mermin,N.D. Phys.Rev.Lett.\textbf{68}
(1992), 557.

\item  Bertrand,J, Calcul des probabilit\'{e}s. Paris: Gauthier-Villars (2nd
ed.1907). Reprinted, New York: Chelsea.,1972.

\item  Bohr,N. Essays 1958-1962 on Atomic Physics and Human Knowledge, Wiley,
New York,1963

\item  Bohr,.N.,Phys.Rev. \textbf{48}(1935), 696 .

\item  Boulant,N.,Havel,T.,Pravia,M.and Cory,D.,Phys.Rev.A\textbf{67}(2003),042322-1

\item  Ballentine, L. E., Rev.Mod.Phys. \textbf{42}(1970)358.

\item  Ballentine, L. E.,Quantum Mechanics: A Modern Development ,World
Scientific , Singapore, 1998

\item  Ballentine, L. E.,Quantum Mechanics: A Modern Development ,World
Scientific , Singapore, 1998, p 238

\item  Bouwmeester,D.,Ekert,A.and Zeilinger,A.,The Physics of Quantum
Information, Springer-Verlag, Berlin, 2000

\item  Bru$\beta,$D., J.Math. Phys.\textbf{43}(2002),\textbf{\ }4237.

\item  Clauser J.F.,Horne M.A.,Shimony A.and Holt R.A.,
Phys.Rev.Lett.\textbf{23}(1969), 880.

\item  Clauser,J.F and Horne,M.A. ,.Phys.Rev.D\textbf{10} (1974),526.

\item  Clauser J.F.and Shimony,A.., Rep.Prog.Phys \textbf{41}(1978) 1883

\item  Einstein, A., in:P.A Schilpp., (ed.), Albert Einstein:
Philosopher-Scientist, Harper \& Row, New York,1949

\item  Einstein.,A., Podolsky B. and Rosen N., Phys.Rev. \textbf{47}(1935), 777.

\item  Ekert,A.,Phys.Rev.Lett. \textbf{67} (1991), 661 .

\item  Gill,R., Time, finite statistics and Bell's fifth position,
arXiv:quant-ph/0312199 (2003).

\item  Golovach, V.N. and Loss D., Semicond.Sci.Technol. \textbf{17} (2002) 355

\item  Grosshans,G., van Assche, G., Wenger,G., Brouri,R., Cerf,N.J. and
Grangier,P, Nature \textbf{421}(2003), 238

\item  Herbert, N.,Am.J.Phys.\textbf{\ 43 } (1975), 315

\item  Ingarden,R.,Kossakowski,A. and Ohya, M., Information Dynamics and Open
Systems,Kluver, Dordrecht, 1997

\item  Jaegger,G. and Sergienko,A. in :E.Wolf,(ed),Progress in
Optics\textbf{\ 42},Elsevier, 2001.

\item  Jamiolkowski,A.,Open.Sys. \& Information.Dyn.\textbf{11}(2004),63

\item  Kielpinski,D.,Monroe,C. and Wineland,D.J.,Nature \textbf{417} (2002), 709.

\item  Kr\"{u}ger,T.,Found.Phys.\textbf{30} (2000),1869

\item  Kupczynski,M. Int.J.Theor.Phys.\textbf{79}(1973), 319, reprinted in:
Physical Theory as Logico-Operational Structure,ed. C.A.Hooker, Reidel,Dordrecht,1978,p.89

\item  Kupczynski,M., New test of completeness of quantum mechanics, ICTP
priprint IC/84/242 (1984)

\item  Kupczynski, M.,Phys.Lett. A \textbf{116}(1986), 417.

\item  Kupczynski,M., Phys.Lett.A \textbf{121}(1987), 51.

\item  Kupczynski,M.,Phys.Lett.A \textbf{121}(1987), 205

\item  Kupczynski, M., On the completeness of quantum mechanics ,
arXiv:quant-ph/028061 ( 2002)

\item  Lewenstein,M., Kraus,.B.,Horodecki,P. and Cirac, J ,Phys.Rev.A
\textbf{63}(2001), 044304

\item  Leonhardt,U., Measuring the quantum states of light, Cambridge
Univ.Press, Cambridge, 1997

\item  Nielsen,M.A. and Chuang.I.L.,Quantum Computation and Quantum
Information,Cambridge Univ.Press, Cambridge, 2000

\item  Rowe,M.A., Kielpinski,D., Meyer,V.,Scakett C.A, Itano, W.M., Monroe,C.
and Wineland,D.J.,Nature \textbf{409}(2001), 791.

\item  Streater,R, J.Math.Phys.\textbf{41}(2000),3556

\item  Weihs,G., Jennewein,T., Simon, C.,Weinfurter,H. and Zeilinger,A.,
Phys.Rev.Lett.\textbf{81}(1998),5039

\item  Wigner,E.P.,Am.J.Phys.\textbf{38}(1970),1003.

\item \.{Z}yczkowski K. and Bengtsson,I., Open.Sys. \&
Information.Dyn.\textbf{11}(2004),3

\item  Allahverdyan A..E,Balian R. and Nieuwenhuizen T.M., The quantum
measurement process in an exactly solvable model, arXiv:cond-mat/0408316 (2004)

\item  Khrennikov,1. A. : ''Interpretation of probability'', VSP, Utrecht, The

Netherlands, (1999).

\item  Holevo,A.S. ''Statistical structure of quantum theory'', - Berlin;

Springer-Verlag, (2001).

\item  Volovich, I.V: ''Quantum cryptography in space and Bell's theorem'', In

the series: PQ-QP: Quantum Probabilility and White Noise Analysis, Volume XIII,

Ed. A. Khrennikov, pp.364-372, World Scientific, (2001).

\item  Khrennikov I. A.and Volovich,I..V: ''Local Realism, Contextualism

and Loopholes in Bell's Experiments'', arXiv: quant-ph/0212127 v1 (2002).

\item  Khrennikov I. A. and Volovich,I.V.: ''Quantum nonlocality, EPR

model and Bell's theorem'', in: 3rd International Sakharov Conference on

Physics. Proceeings, v.II, eds. A. Semikhatov, M. Vasiliev and V. Zaikin,

pp.260-267, World Scientific, (2003).
\end{enumerate}
\end{document}